\documentclass[pra,amsfonts,showpacs]{revtex4}

\usepackage[dvips]{graphicx}
\usepackage{bm}

\renewcommand{\deg}{$^{\circ}$}

\begin{document}

\title{The Stoner-Wohlfarth model of Ferromagnetism: Static properties}

\author{C. Tannous and J. Gieraltowski}
\affiliation{Laboratoire de Magn\'etisme de Bretagne - CNRS FRE 2697\\
Universit\'e de Bretagne Occidentale -\\
6, Avenue le Gorgeu C.S. 93837 - 29238 Brest Cedex 3 - FRANCE}

\begin{abstract}
Recent advances in high-density magnetic storage and spin electronics are
based on the combined use of magnetic materials with conventional microelectronic
materials (metals, insulators and semiconductors). The unit of
information (bit) is stored  as a magnetization state in some ferromagnetic material (FM)
and controlled with an external field altering the magnetization state. 
As device size is shrinking steadily toward the nanometer and the need to increase the processing 
bandwidth  prevails, racing toward higher frequencies is getting even more challenging.
In magnetic systems,  denser storage leads to  finer magnetic grains and smaller size
leads to single magnetic domain physics. 
The Stoner-Wohlfarth model is the simplest model that describes adequately the 
physics of fine magnetic grains containing single domains and where  
magnetization state changes by rotation or switching (abrupt reversal). 
The SW model is reviewed and discussed with its consequences and potential applications 
in the physics of magnetism and spin electronics. 

\end{abstract}

\pacs{51.60.+a, 74.25.Ha, 75.00.00, 75.60.Ej, 75.75.+a}

\keywords{Magnetic properties. Magnetic materials. Hysteresis. 
Magnetic properties of nanostructures }

\maketitle

\section{Introduction}
Moore's law of Microelectronics (currently still valid since more than 40 years)
states that one should expect a doubling of CPU performance every 18 to 24 months. 
While high-density magnetic storage is advancing (over the long term) at a rate
similar to Moore's law, rates surpassing Moore's law were recently observed. 

When bit density $d$ gets larger, bit length that scales as $\sim 1/\sqrt{d}$
gets smaller at a rate that recently reached almost the double value 
of the microelectronics feature size rate.
In the latter case, a bit is stored as an electronic charge as in Flash memory, 
flip-flops, registers and cache memory (the charge can also be periodically refreshed as in 
capacitor based random access memories D-RAM),  while in mass storage magnetic 
media (Disk, floppy, tape...) a bit corresponds to a well defined orientation of magnetization and is
usually stored within many ferromagnetic grains possessing an average orientation.

When the typical  bit size shrinks, we get closer to the limit where one magnetic 
grain is able of holding a single bit of information.
This is the realm of nanometer electronics (nanoelectronics) and spin electronics (spintronics).
Spintronics is a new field of electronics with the coexistence
of classical materials used in standard electronics (metals, semiconductors and
insulators) and magnetic materials (ferromagnetic, anti-ferromagnetic, ferrimagnetic,
paramagnetic, diamagnetic etc...).

The goal of spintronics~\cite{zutic} is to make novel devices that are controlled by the combined actions
of electric and magnetic fields.  A spintronic device has an additional 
degree of freedom with respect to a standard device controlled by charge only; 
it has also a polarization (or a magnetic state: up or down in the transverse
technology and right left in the longitudinal case). Transverse and longitudinal 
technologies refer to perpendicular or parallel to the direction
set by the media mechanical rotation velocity (Disk or floppy). \\
In an ordinary p-n junction diode, the total current is obtained 
from electronic $n$ and hole $p$ charge densities, whereas in a spin polarised diode, 
the current is composed of $n_{\uparrow}$, $n_{\downarrow}$ and
$p_{\uparrow}$, $p_{\downarrow}$; hence spin degrees of freedom 
${\uparrow}, {\downarrow}$ play a role in addition to charge. \\
A magnetic memory such as an MRAM retains any information stored 
 (being of magnetic type)
even when the system is powered down (or hangs up) in contrast to electronic RAM's 
and registers (since information is of electronic type). 
That means, in a PC containing MRAM's the operating system is
loaded once for all (at first boot) and if the system hangs or is powered down, all
temporary information is preserved. Additionally the magnetic polarization 
degree of freedom can lead to new devices of interest in quantum 
computing, communication and storage devices, so called quantum information
devices.  \\

The Stoner-Wohlfarth (SW) model of Ferromagnetism is the simplest model
that is adequate to describe the physics of tiny magnetic grains containing
single magnetic domains.
It can be considered as a sort of Hydrogen model
of Ferromagnetism. The physics of the SW model is built on a series
of assumptions that ought to be placed into perspective in order to highlight 
and understand the progress and insight in magnetism and magnetic materials. \\

This Redux is made of two parts: the first tackles the static
properties and the second the dynamic and statistical properties 
(dynamics of the magnetization reversal under either temperature or a time 
dependent magnetic field). \\
This part is organised as follows: 
section 2 describes the basics of the Stoner-
Wohlfarth model. Section 3 describes the hysteresis curves associated with the model.
Section 4 details the energetics of the SW model (nature of the energy barrier to
cross when there is a change in magnetization state).
Finally, section 5 discusses some of the limitations of the model.

\section{The Stoner-Wohlfarth model}
\begin{figure}[!h]
\begin{center}
\scalebox{1.2}{\includegraphics*{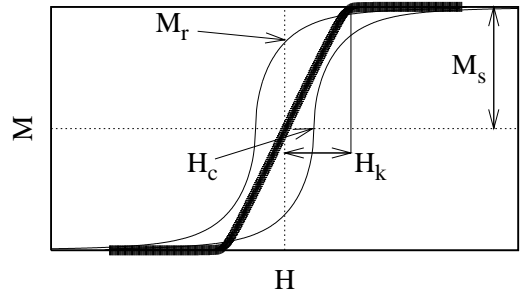}}
\end{center}
  \caption{Single domain hysteresis loop obtained for an arbitrary angle,  $\phi$,  between the magnetic
 field and the anisotropy axis~\cite{ea}. Associated quantities such as coercive
field $H_{c}$, anisotropy field $H_{K}$ and remanent magnetization $M_r$ are shown. The thick line
is the hysteresis loop when the field is along the hard axis ($\phi$=90 degrees,  in this case) and
$H_{K}$ is the field value at the slope break. Quantities such as $H_{c}$ and $M_r$
 depend on $\phi$ whereas the saturation magnetization $M_s$ does not.}
\label{hyster}
\end{figure}

A FM used typically in recording media (hard disks, floppies,
tapes \ldots) can be considered as made of a large number 
 of interacting magnetic moments (on the order of $N_A$ the Avogadro number for
a mole of material).  \\
When a magnetic field is applied to a FM, 
a magnetization change takes place. One way to understand the 
underlying phenomena is to plot the value of the magnetization $\bm{M}$ 
projected along the direction of the applied field $\bm{H}$. The locus of 
the magnetization $\bm{M}$ measured along the direction of the applied 
magnetic field $\bm{M(H)}$ versus the field and depicted in the 
$\bm{M-H}$ plane is the hysteresis loop (see fig.~\ref{hyster}). 
The term hysteresis (delay in Greek) means that 
when the material is field cycled (i.e. the field $\bm{H}$ is increased 
then decreased) two different non-overlapping curves ("ascending and 
descending branches") $\bm{M(H)}$ are  obtained.

\begin{figure}[!h]
\begin{center}
\scalebox{0.5}{\includegraphics*{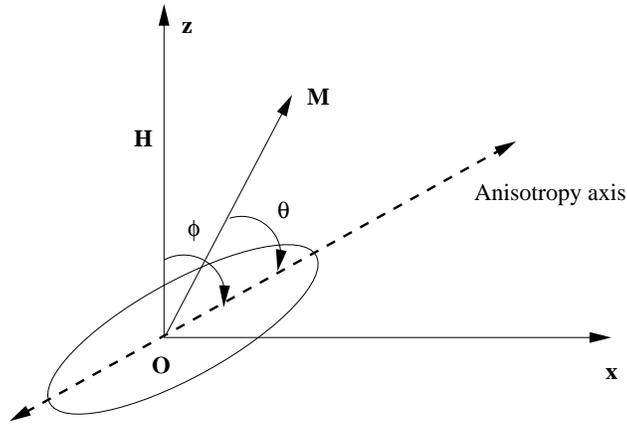}}
\end{center}
  \caption{Single domain grain with the magnetization $\bm{M}$ and the external applied
magnetic field. The anisotropy of strength $K$ competes with the magnetic field $\bm{H}$ along
the $z$ axis in the alignment of the magnetic moment that is restricted to the 2D $xOz$ plane.
Note that the anisotropy energy~\cite{energy} $K_{ij}\bm{M}_{i}\bm{M}_{j}/M_s^2$ reduces in the uniaxial case 
to $K\sin ^2\theta$. It is minimum (=0) when $\theta=0$ and $K>0$ (see note~\cite{note}.}
\label{grain}
\end{figure}

The main characteristics of the hysteresis loop are
the saturation magnetization $M_{s}$ (saturation is attained when all
the magnetic moments are aligned along some common direction resulting
in the largest value of the magnetization), the remanent magnetization $M_r$ 
(the leftover magnetization i.e. magnetization when the field $H=0$ is
the basis of zero-cost information storage with $\pm M_r$ orientations representing
a single bit.) and the coercive field $H_{c}$ (which makes $\bm{M=0}$, i.e. the field
such that $\bm{M(H_c)=0}$) and the anisotropy field 
$H_{K}$. The hysteresis loop might be viewed 
as a sort of magnetic $I-V$ characteristic ($I$ corresponds to $M$ and 
$V$ to $H$). The characteristic is non-linear and the output $M$ is delayed with respect 
to input $H$. The input-output delay is proportional to the width of the loop.
The ratio $M_{r} /M_{s}$ called squareness is close to 1 when the applied 
magnetic field is close to some orientation defined as the easy axis (EA~\cite{ea}) and  the 
hysteresis loop is closest to a square shape.
Once the EA is determined, the angle the magnetic field makes with the EA (say 
$\phi$) is varied and the hysteresis loop is graphed for different angles (see fig.~\ref{grain}). 
When the angle $\phi $ is increased the opening  
of the hysteresis loop is reduced; it is largest when the magnetic field is most parallel to the  
EA and smallest when the magnetic field is most parallel to the so called 
hard axis (in simple systems the hard axis is perpendicular to the EA).
As shown in fig.~\ref{hyster}, most characteristics of the 
hysteresis loop are depicted and for a given temperature and frequency of 
the applied field $H$, quantities such as the remanent magnetization $M_{r}$, 
and the coercive field $H_{c}$ depend on the angle $\phi $. \\

When temperature or field frequency are 
varied, the hysteresis loop shape changes and may even be seriously altered by the 
frequency of the field. The  hysteresis loop branches might collapse altogether
over a single curve  above a given temperature (Curie temperature), the material becoming
paramagnetic and therefore unable to store information. \\

The SW model (also called coherent rotation~\cite{Chikazumi}) considers a FM
 as represented by a single magnetic moment
(thus the name coherent as in optics for fixed phase relationship or as in a
superconductor where a single wavefunction represents all electrons in the material). 
The material is therefore considered as a single magnetic
domain, thus all domain related effects  or inhomogeneities are not considered.
A single domain occurs when the size  of the grain is smaller than some length termed
the critical radius (see subsection II-A) and  contains about 
10$^{12}$-10$^{18}$ atoms, typically.\\ 
At $T=0K$ a grain carrying a single moment $\bm{M}$, is an ellipsoid-shaped object (see fig.~\ref{grain}) 
since a material with uniform magnetization ought to have an ellipsoid form \cite{landau}. \\
The grain possesses a uniaxial anisotropy (meaning an axis along which
the magnetization prefers to lie in order to minimize the energy) and is subjected 
to an externally applied static magnetic field $\bm{H}$.
$\bm{M}$ evolves strictly in a two dimensional space  (see fig.~\ref{grain}), therefore it is 
characterized by a single angle $\theta$, the angle $\bm{M}$ makes with the anisotropy axis
(also the EA, in this case~\cite{ea}). $\phi$  is the angle 
the external applied magnetic field (taken as the $z$-axis) makes with the EA. \\

Thus, the moment $\bm{M}$ is subjected to two competing 
alignment forces: one is due to a uniaxial anisotropy characterised by $K$ favoring some 
direction (see fig.~\ref{grain}) and the other is due to an external magnetic field $\bm{H}$. 
Therefore, the total energy is 
the anisotropy energy $E_{A}$ and the Zeeman energy $E_{Z}=-\bm{M} \cdot \bm{H}$.
At $T=0K$, the energy (per unit volume~\cite{energy}) is then:
\begin{equation}
E = E_A + E_Z = K\sin ^2\theta - H M_s \cos ( \theta - \phi)
\end{equation}
The anisotropy energy~\cite{energy} $K\sin ^2\theta$ is minimum (=0) when $\theta=0$
for $K>0$ (see note~\cite{note}). 
The moment will select a direction such that the total energy $E_{A} + E_{Z}$ is 
minimized. The orientation change might occur smoothly (rotation)  or suddenly (switching)
implying that the magnetization is
discontinous at some value of the magnetic field $\bm{H}$. \\
A FM of a finite size (like the ellipsoidal grain) magnetized uniformly 
(the magnetization $\bm{M}$ is represented by its components $\bm{M}_{\alpha })$ 
contains a magnetic energy (called also magnetostatic energy~\cite{energy}) given by 
$2\pi {N}_{ij}\bm{M}_{i }\bm{M}_{j }$ 
(Einstein summation is considered). The $N_{ij }$ coefficients are the 
demagnetization coefficients of the body determined by its shape. 
The origin of the terminology is due to the resemblance to the familiar 
anisotropy energy~\cite{energy} of the form $K_{ij}\bm{M}_{i}\bm{M}_{j}/M_s^2$. 
The coefficients depend on the geometry 
of the material. For simple symmetric geometries, one has three positive coefficients along three 
directions $N_{xx}$, $N_{yy}$ and $N_{zz}$ (the off-diagonal terms 
are all 0). This is the case of wires, disks, thin films and 
spheres. All three coefficients are positive, smaller than 1 and their 
sum is equal to 1. For a sphere, all three coefficients are equal to  $\frac{1}{3}$. For 
a disk they are given by 0,0,1 if the $z$ axis is 
perpendicular  the disk lying in the $xy$ plane. For an 
infinite length cylindrical wire with its axis lying along the $z$ direction, 
the values are $\frac{1}{2}, \frac{1}{2},0$~\cite{Chikazumi}. \\

A FM (assumed to possess a single uniform magnetization $M$ even though it 
does not have an ellipsoid shape) cut in the form of a cylinder, disk or a 
thin film possessing a uniaxial anisotropy can be considered as having 
the anisotropy energy $K\sin^{2}\theta$ with $K>0$ (see \cite{note}). If we add to anisotropy the 
demagnetization energy $2\pi {N}_{\alpha \beta }\bm{ M}_{\alpha }\bm{M}_{\beta }$
(due to the body finite size along one or several directions), we get a competition between 
the two energies:
\begin{equation}
K\sin ^2\theta + 2\pi  \,(N_ \bot M_x^2 + N_ \bot M_y^2 + N_\parallel M_z^2 ) 
\end{equation}
This yields: 
\begin{equation} 
\left[ K + 2\pi  M_s^2 \,(N_ \bot - N_\parallel ) \right] \sin ^2\theta + \mbox{  const.}
\end{equation}

The indices $\parallel, \bot $ of the 
demagnetization coefficients denote respectively parallel or perpendicular to the $z$ axis. 
Thus we define an effective anisotropy $K_{eff}$ as:

\begin{equation}
K_{eff} = [K + 2\pi  M_s^2 \,(N_ \bot - N_\parallel )]\sin^2\theta 
\end{equation}

The total energy~\cite{energy} of the SW model (called from now on Stoner particle) can then be written as:
\begin{equation}
E=K_{eff} \sin^2(\theta) -M_s H \cos(\theta-\phi)
\end{equation}

\begin{figure}[!h]
\begin{center}
\scalebox{0.3}{\includegraphics*[angle=-90]{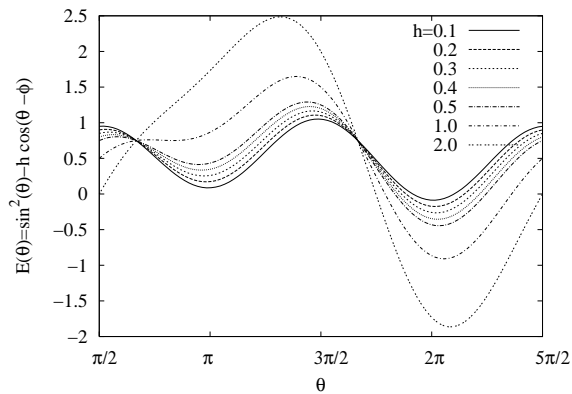}}
\end{center}
  \caption{Variation of the  energy~\cite{energy} landscape with angle $\theta$ for different values of the 
normalised magnetic field $h=H/H_K$.  The field orientation is held fixed at the value
 $\phi=30$\deg with the anisotropy axis.}
\label{energy}
\end{figure}

At equilibrium, the magnetization points along a direction defined by an angle
 $\theta^*$ that minimizes the energy.
The behaviour of the energy as a function of $\theta$ for a fixed angle 
$\phi=30$\deg  and for various fields $h=H/H_K$ (the anisotropy field $H_K=2K_{eff}/M_s$) 
is depicted in fig.~\ref{energy}. One observes that
despite the wide variability of the energy landscape versus $\theta$, a couple of points are
 not affected by $h$. In addition, the landscape becomes quite flat for some value of $h$.\\
The minimum condition at $\theta^*$ is:
\begin{equation}
\left( {\frac{\partial E}{\partial \theta }} \right)_{\theta=\theta^*} =0 \hspace{0.5cm} \mbox{and} \hspace{0.5cm}
\left( {\frac{\partial ^2E}{\partial \theta ^2}} \right)_{\theta=\theta^*} > 0
\label{eq:min0}
\end{equation}

Normalizing the magnetization by its saturation value, such that $m=M/M_s$,  yields:

\begin{eqnarray}
\left[ \sin(\theta)\cos(\theta)+h \sin(\theta-\phi) \right]_{\theta=\theta^*}  & = 0  \\ 
 \mbox{and} \hspace{0.5cm} \left[  \cos(2\theta)+h \cos(\theta-\phi) \right]_{\theta=\theta^*} & \ge 0
\label{eq:min}
\end{eqnarray}

For general $\phi$, the above equations cannot be solved analytically, except for $\phi=0, \pi/4, \pi/2$.\\
In order to present the possible solutions, we define two components of the magnetization: 
\begin{enumerate}
\item The longitudinal magnetization i.e. the projection of $\bm{M}$ along $\bm{H}$, 
$m_{\|}=\cos(\theta-\phi)$.

\item The transverse magnetization i.e. the projection of $\bm{M}$ perpendicularly to $\bm{H}$, 
$m_{\bot}=\sin(\theta-\phi)$.
\end{enumerate}

\begin{figure}[!h]
\begin{center}
\scalebox{0.3}{\includegraphics*{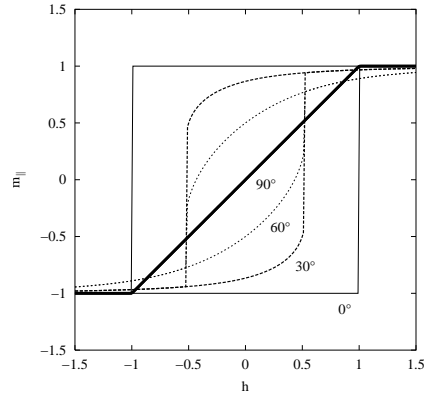}}
\end{center}
  \caption{Longitudinal hysteresis loop for various angles $\phi$ of the field $H$ with the EA. Note the 
square loop for $\phi=0$ and the diagonal line $m_{||}=h$ for $\phi=\pi/2$.}
\label{vsm}
\end{figure}

\begin{figure}[!h]
\begin{center}
\scalebox{0.3}{\includegraphics*{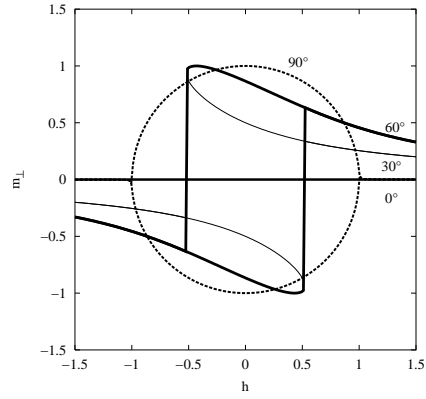}}
\end{center}
  \caption{Transverse hysteresis loop for various angles $\phi$ of the field $H$ with the EA. Note the 
horizontal line $m_\bot=0$ for $\phi=0$ and the circle for $\phi=\pi/2$.}
\label{rtm}
\end{figure}

Let us find the minimum analytically for the cases: $\phi=0 \mbox{  and  }  \pi/2$.
\begin{enumerate}
\item $\phi=0$: eqs.~\ref{eq:min} give the solution: \\
$ \theta^{*}=\cos^{-1}(-h) \mbox{  when   } h \le 1  \mbox{  otherwise   }  \theta^{*}=0, \pi$
 yielding the square hysteresis loop in fig.~\ref{vsm} and the line $m_{\bot}=0$ in
fig.~\ref{rtm}.

\item $\phi=\pi/2$: eqs.~\ref{eq:min} give: \\
$ \theta^{*}=\sin^{-1}(h) \mbox{  when   } h \le 1  \mbox{  otherwise   }  \theta^{*}=\pi/2$
 yielding the main diagonal ($m_{\|}=h$) hysteresis loop in fig.~\ref{vsm} 
and the circle $m_{\|}=\pm \sqrt{1-h^2}$ in fig.~\ref{rtm}.

\end{enumerate}

Remarkably, both types of hysteresis curves (depicted in fig.~\ref{vsm} and fig.~\ref{rtm})
exist and are encountered in many physical systems. \\
The questions that arise are then:
\begin{itemize} 
\item What are the conditions for expecting a  Stoner particle behaviour?
\item Is there some characteristic size below which a single domain is expected?
\end{itemize}

Attempt at answering the above are discussed next.

\subsection{Stoner particle and critical radius}
\begin{figure}[!h]
\begin{center}
\scalebox{0.5}{\includegraphics*{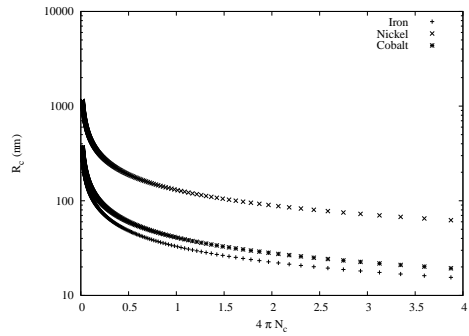}}
\end{center}
  \caption{Critical radius (nm) or single domain size for Iron, Nickel and Cobalt for a prolate 
ellipsoid (elongated) as  a function of the demagnetization coefficient along its axis $N_c$. 
The room-temperature data for these curves (taken from Kittel~\cite{Kittel}) are $M_s=$ 1710 (Fe), 485 (Ni) and
 1440 (Co) (in Gauss); $a_1=$ 2.48 (Fe), 2.49  (Ni) and 2.50  (Co) (in \AA).
 The exchange stiffness constant $A$ in all cases is taken as 10$^{-6}$ erg/cm}
\label{radius}
\end{figure}

When a magnetic medium is made of non-interacting grains, there is possibility for
observing Stoner particle behaviour (single domain) when the typical size of the 
grain is below the critical radius $R_c$ of the grain.
A simple argument given in Landau-Lifshitz {\it Electrodynamics of Continuous Media}~\cite{landau}
 is based on the following: When the demagnetization
energy~\cite{energy} $2\pi {N}_{ij}\bm{M}_{i }\bm{M}_{j } \sim 2\pi {N}_{c} M_s^2$ (where ${N}_{c}$ is the
demagnetization coefficient along some preferred axis, usually the long one in an ellipsoid-shaped grain) is equal to the 
exchange energy~\cite{energy} 
$\frac{A_{ij}}{M_s^2} \frac{\partial M_k}{\partial x_i} \frac{\partial M_k}{\partial x_j} \sim \frac{A}{R_c^2}$
($A_{ij}$, $i,j,k=1,2,3$ is the exchange stiffness constant along $i,j$ directions). \\
Considering that $A_{ij} \sim A$ a typical exchange stiffness constant (regardless of $i,j$)
results in $R_c \sim \sqrt{\frac{A}{2\pi N_c M_s^2}}$. Exchange energy is the largest contribution
to non-uniformity energy due to spatial variation of the magnetization $\bm{M}$.\\
 When the change in the direction
of $\bm{M}$ occurs over distances that are large compared to interatomic distances, non-uniformity energy 
can be expressed with derivatives of $\bm{M}$ with respect to spatial coordinates (see Landau-Lifshitz~\cite{landau}
 and Brown~\cite{Brown}). Exchange stiffness constant $A_{ij}$ is on the order of Heisenberg exchange energy 
per unit length $J/a_1$  ($a_1$ is the average nearest neighbour distance within the grain). 
Typically $J \sim 10$ meV and $a \sim $ 1\AA, hence
we get $A_{ij} \sim $  10$^{-6}$ erg/cm (see fig.~\ref{radius}). This length is in fact on
the order of the domain wall thickness, therefore we rather rely on 
Frei et al.~\cite{frei} approach to estimate the critical radius.
They define  $R_c$ from a minimization
of the energy using Euler variational equations obtaining the equation:

\begin{equation}
R_c-\sqrt{\frac{3 A}{2 \pi N_c M_s^2} [ \ln(\frac{4 R_c}{a_1})-1]}=0
\label{eq:rc}
\end{equation}

Solving eq.~\ref{eq:rc} for $R_c$, in the case of standard ferromagnetic transition metals  
(Fe, Ni and Co), fig.~\ref{radius} gives the variation of $R_c$ with $N_c$ 
(the long axis demagnetization coefficient). 
From the figure, we infer that for elongated 
grains made of Iron, Nickel or Cobalt, $R_c$ is within a few 100 nm range.

\section{Hysteresis in the Stoner-Wohlfarth model}

Applying a magnetic  field $\bm{H}$ to a FM and measuring a resulting magnetization 
$\bm{M}$ as a response can be considered as a classical signal input-output problem.
The input-output characteristic $\bm{M(H)}$ is that of a peculiar 
non-linear filter except at very low fields where
$\bm{M}$ is simply proportional to $\bm{H}$. A 
simple illustration of non-linearity is to observe the output as a square 
signal whereas the input is a sinusoidal excitation (see ref.~\cite{chakra}).
In addition, the material imposes a propagation delay to the signal 
proportional to the width of the hysteresis loop (twice the coercive field). \\
Hysteretic behaviour is generally exploited in control systems, for instance,
because different values of the output are required 
as the input excitation is varied in an increasing or decreasing fashion. 

We consider two types of hysteresis curves, as mentioned in the previous section:
\begin{enumerate}
\item A longitudinal hysteresis curve with the reduced magnetization $m_{\|}$ taken
along the direction of the applied external magnetic field (see fig.~\ref{vsm}).
\item A transverse hysteresis curve with the reduced magnetization  $m_{\bot}$ taken
along the direction perpendicular to the applied external field (see fig.~\ref{rtm}).
\end{enumerate}

As an illustration, the hysteresis loop depicted in fig.~\ref{hyster}, is found by calculating the 
component of $\bm{M}$ along $\bm{H}$ from the set of $\theta $ angles at 
a given angle $\phi $, that minimize the energy $E$ (conditions given in eq.~\ref{eq:min}).
The anisotropy field $H_{K}$  is obtained from the slope break of the hysteresis 
loop when $\phi =\pi/2$, that is when $\bm{H}$ is along the hard axis. 
In this simple model, the coercive field $H_{c}$ (at $\phi $=0) (for which 
$M=0$) is found as $H_{c}= H_{K}$. The loop is found to be broadest when 
$\phi $=0, and it gets thinner as $\phi $ is increased to collapse into a simple 
line (for $\phi =\pi/2$). That line breaks its slope in order to reach saturation behaviour
($M=\pm M_{s})$ for $H= \pm H_{K}$ and the coercive field is zero in that case (see fig.~\ref{critHs}).\\

The question of the occurrence of hysteresis is addressed next.

\subsection{Magnetization reversal in the Stoner-Wohlfarth model}

\begin{figure}[!h]
\begin{center}
\scalebox{0.5}{\includegraphics*[angle=0]{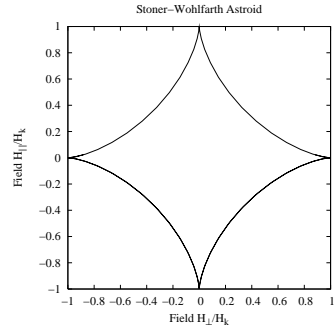}}
\end{center}
  \caption{The inside the astroid domain is made of the field values for which a reversal of the
magnetization is possible. Outside the astroid domain, no reversal is possible.}
\label{astroid}
\end{figure}

Hysteresis boundaries versus  applied field are determined from the simultaneous
nulling of the first and second derivative of the energy (that refers to the observed
flatness of the energy landscape versus $\theta$ as discussed previously). 
Thus one obtains the astroid equation (see fig.~\ref{astroid}):

\begin{equation}
\left( {\frac{H_\bot }{H_K }} \right)^{2 / 3} + \left( {\frac{H_{\parallel} }{H_K }} 
\right)^{2 / 3} = 1
\end{equation}

The fields $H_{\bot}$ and $H_{\parallel}$ are the components of the field $\bm{H}$ 
along the hard and easy axes. The critical field (equal in this case to the 
coercive field at $\phi $=0) for which $M$ jumps (at a given orientation of 
the field) is obtained from the conditions:

\begin{equation}
\left( {\frac{\partial E}{\partial \theta }} \right)_\phi = 0 \hspace{1cm} \mbox{and} \hspace{1cm}
\left( {\frac{\partial ^2E}{\partial \theta ^2}} \right)_\phi = 0
\label{eq:switching}
\end{equation}

as:

\begin{equation}
H_s (\phi ) = \frac{H_K }{\left[ {\sin ^{2 / 3}\phi + \cos ^{2 / 3}\phi } 
\right]^{^{3 / 2}}}
\end{equation}

In spite of the tremendous simplifying assumptions of the SW model and 
the fact several derived quantities appear to be equal (e.g. the coercive 
field at $\phi $=0 and the anisotropy field $H_{K}$), it is extremely helpful since 
it captures, in many cases, the essential physics of the problem; in 
addition, many quantities of interest can be derived analytically (see 
ref.~\cite{stoner}).\\

\begin{figure}[!h]
\begin{center}
\scalebox{0.3}{\includegraphics*[angle=-90]{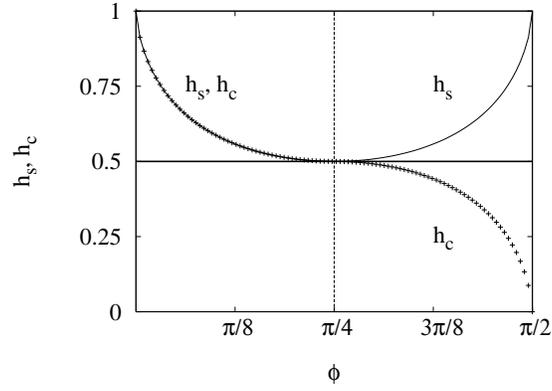}}
\end{center}
  \caption{Normalised switching $h_s=H_s/H_K$ and coercive $h_c=H_c/H_K$  fields versus angle $\phi$ showing 
that the minimum field to reverse the magnetization is half
the anisotropy field. In fact, this is valid in the static case only. In the dynamic case where the applied
field is time dependent the minimum field can be quite smaller as discussed in the next part of the paper.
The coercive field is equal to the switching field for $\phi < \pi/4$ and to its mirror with 
respect to the horizontal line 1/2 for $\phi > \pi/4$.}
\label{critHs}
\end{figure}

The critical field (called henceforth switching field)  for which the magnetization value 
jumps from one energy minimum to another equivalent to the first is denoted $H_s$. 
We use the $s$ index referring to switching in order to distinguish $H_s$
from the coercive field $H_c$ (see fig.~\ref{critHs}). \\
Depending on the shape of the hysteresis loop, there are two cases to consider:
\begin{enumerate}

\item Derivative case: $H_s$ may be considered as an external magnetic field  for which the 
absolute derivative $\vert dM/dH \vert $ diverges or is very large, meaning 
that we have ${\vert dM/dH \vert}_{H = H_s } \rightarrow \infty $.

\item Energy case: $H_s$ is found below from an energy equality condition: 
$[E(M_1 )]_{H = H_s } = [E(M_2 )]_{H = H_s }$ 
where $E(M_{1})$ (resp. $E(M_{2}))$ is the energy with $M_{1,2}$ the magnetization 
in the first minimum state (resp. in the second state).
\end{enumerate}

The above conditions insure the magnetization jumps from one minimum to another.
The nulling of energy angle derivatives (first and second) yields a flat energy
behaviour versus angle (see eq.~\ref{eq:switching}) facilitating such
jumping (see note~\cite{jump}). 

In  fig.~\ref{critHs} the switching field $H_s$ versus angle $\phi$ is
displayed showing the minimum field to reverse the magnetization, being half
the anisotropy field, must be applied with an angle of 135 \deg (90\deg +45\deg) 
with respect to the EA. 

\begin{figure}[!h]
\begin{center}
\scalebox{0.3}{\includegraphics*[angle=-90]{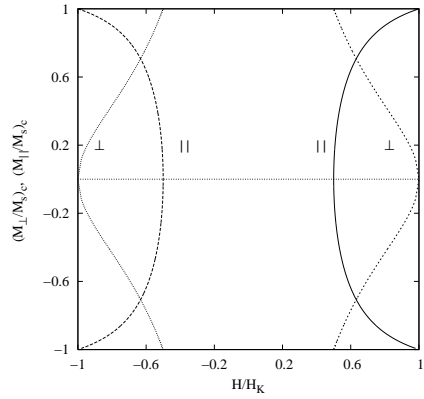}}
\end{center}
\caption{Critical longitudinal  $m_{\|,c}=\cos(\theta_c-\phi)$
labeled as $\|$ and transverse magnetization $m_{\bot,c}=\sin(\theta_c-\phi)$
labeled as $\bot$ versus magnetic field.}
\label{critLT}
\end{figure}

The critical magnetization is the magnetization at $H_s$ and can be evaluated from the critical
angle: $\theta_c= \phi+ \tan^{-1}{[\tan(\phi)]}^{1/3}$.
We use this value to follow the variation of the longitudinal $m_{\|,c}=\cos(\theta_c-\phi)$
and transverse magnetization $m_{\bot,c}=\sin(\theta_c-\phi)$
as functions of the switching field as displayed in fig.~\ref{critLT}.

\section{Barrier height and its dependence on applied field}

It is interesting to examine the behaviour of the energy barrier separating
two magnetization states with an applied field. A deep understanding of the 
nature of the barrier, its characteristics and how it is altered by
material composition, field or temperature will help us control and finely tune 
the behaviour of magnetic states inside a FM.

Some of the questions one might ask are the following:
\begin{enumerate}
\item What are the exact characteristics of the energy barrier $\Delta E$?
\item Does $\Delta E$ vary with the nature and shape of the magnetic grain?
\item How does $\Delta E$ vary with an external applied field?
\item If there is some interaction between grains, how does it affect the barrier height?
\end{enumerate}

\begin{figure}[!h]
\begin{center}
\scalebox{0.3}{\includegraphics*[angle=-90]{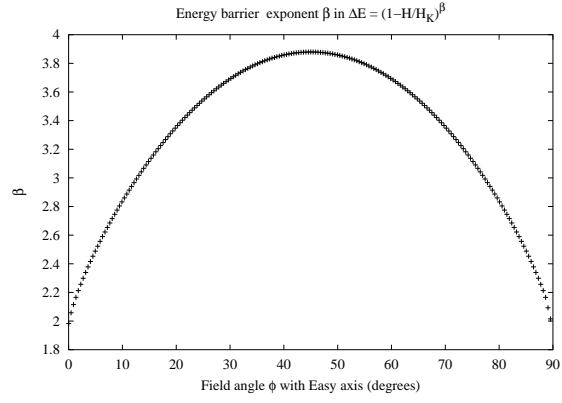}}
\end{center}
  \caption{Variation of the barrier exponent $\beta$ with field normalised with the anisotropy field.
The barrier is given by $\Delta E={(1-H/H_K)}^\beta$.}
\label{expohk}
\end{figure}

\begin{figure}[!h]
\begin{center}
\scalebox{0.3}{\includegraphics*[angle=-90]{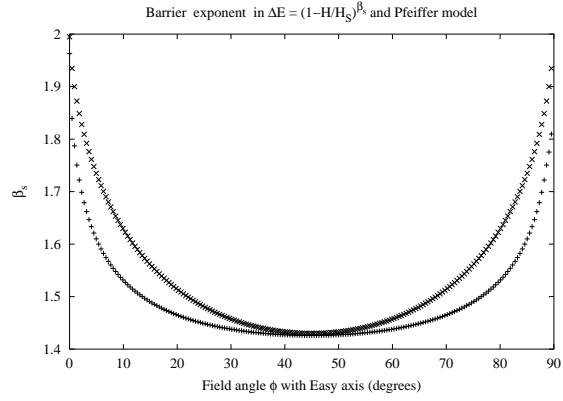}}
\end{center}
  \caption{The lower curve depicts the barrier exponent $\beta_s$ variation with field normalised
 with respect to the switching field.
The barrier is given by $\Delta E={(1-H/H_s)}^{\beta_s}$. The upper curve is 
 the Pfeiffer approximation for the exponent with the barrier expression given by 
$\Delta E= {(1-H/H_s(\phi))}^{[0.86+1.14 H_s(\phi)]}$.}
\label{expohs}
\end{figure}

First of all, the  energy barrier is defined as the minimum energy separating 
two neighbouring energy minima (one of them being a local minimum).
The applied field modifies the shape of the energy barrier as depicted in 
fig.~\ref{energy} leading to consider at least two field choices ($H_K$ or $H_s$), i.e:
$\Delta E={(1-H/H_K)}^\beta$ or $\Delta E={(1-H/H_s)}^{\beta_s}$ where $\Delta E$ is normalised
by the effective anisotropy constant.
The behaviour of the barriers are depicted in fig.~\ref{expohk} and fig.~\ref{expohs}. 
The often quoted exponent $\beta=2$ is valid only when $\phi=0, \mbox{ or }  \pi/2$. 
Moreover we observe that when we normalize the applied field with  respect to
$H_K$ (fig.~\ref{expohk}) the curvature is opposite to  what is observed in the 
$H_s$ normalization case (fig.~\ref{expohs}). 
Again the often cited exponent 2 is valid only when $\phi=0, \mbox{ or }  \pi/2$
exactly as in the $H_K$ normalization case. \\ 
A popular analytic approximation for the barrier is the Pfeiffer approximation given by the formula:
$\Delta E= {(1-H/H_s(\phi))}^{[0.86+1.14 H_s(\phi)]}$ yielding an exponent
$\beta_s = 0.86+1.14 H_s(\phi)$. Fig.~\ref{expohs} indicates that this
approximation is not so bad when compared to our exact numerical
calculation.\\
When an assembly of interacting grains are considered, it is possible
to recast the barrier formula in a form similar to the non-interacting Stoner particle case,
however, the nature of the interaction between the grains and the way it is accounted 
for will determine the value of the exponent.

\section{Limitations of the Stoner-Wohlfarth model}
The SW model is a macrospin approach to magnetic systems (like "coarse-graining" 
in Statistical physics, or Ehrenfest approach in Quantum systems) because of the complexity
of a direct microscopic (nanoscopic being more appropriate) description. 

Its main concern is single domain physics despite the fact magnetic
materials, in general, possess a multi-domain structure.
Regardless of this consideration, several limitations are already built into the SW model.
One limitation is the crossover problem indicating that hysteresis branches may 
cross for certain values of the angle $\phi$.

The minimum energy equation may be cast in the form:
$\sin(\theta)\cos(\theta)+h \sin(\theta-\phi)=0$ through the replacement:
$m=\cos(\theta-\phi)$ obtaining the expressions for 
the upper and lower branches: 

\begin{eqnarray}
h_{\uparrow}  & =-m\cos(2\phi) + \frac{(2m^2-1)}{2\sqrt{1-m^2}}\sin(2\phi)  \\
h_{\downarrow} & =-m\cos(2\phi) - \frac{(2m^2-1)}{2\sqrt{1-m^2}}\sin(2\phi)
\end{eqnarray}

A crossing (crossover) between the two branches occurs at $h_x$ 
when $h_{\uparrow}= h_{\downarrow}$. Solving this
equation, we obtain $m=\pm 1/\sqrt{2}$ yielding $h_{x}=-\cos(2\phi)/\sqrt{2}$. The crossing
angle is defined as the angle for which the field $h_x$ is equal to
the switching field, that is for an angle $\phi_x$ satisfying:

\begin{equation}
\cos 2 \phi_x= -\frac{\sqrt{2}}{{[\sin ^{2 / 3}(\phi_x) + \cos ^{2 / 3}(\phi_x) ]}^{^{3 / 2}}}
\end{equation}

Writing $u=\cos^2 \phi_x$ we transform this equation into:
$ {(1-u)}^{1/3}+{u}^{1/3}= \frac{{2}^{1/3}}{ {(1-2u)}^{2/3}}$.
Using the transformation, $u=\frac{1}{2}(1-\frac{2}{\sqrt{v}})$ 
we finally obtain the equation in $v$ as:

\begin{equation}
{[1+\frac{2}{\sqrt{v}}]}^{1/3} + {[1-\frac{2}{\sqrt{v}}]}^{1/3}= v^{1/3},  \mbox{     } v > 0
\end{equation}

Remarkably, this transcendental equation has a unique  integer solution $v=5$. \\
Hence the crossover angle $\phi_x=\frac{1}{2} \cos^{-1}( -\frac{2}{\sqrt{5}}) \sim 76.72 \mbox{\deg}$.

\begin{figure}[!ht]
\begin{center}
\scalebox{0.3}{\includegraphics*[angle=-90]{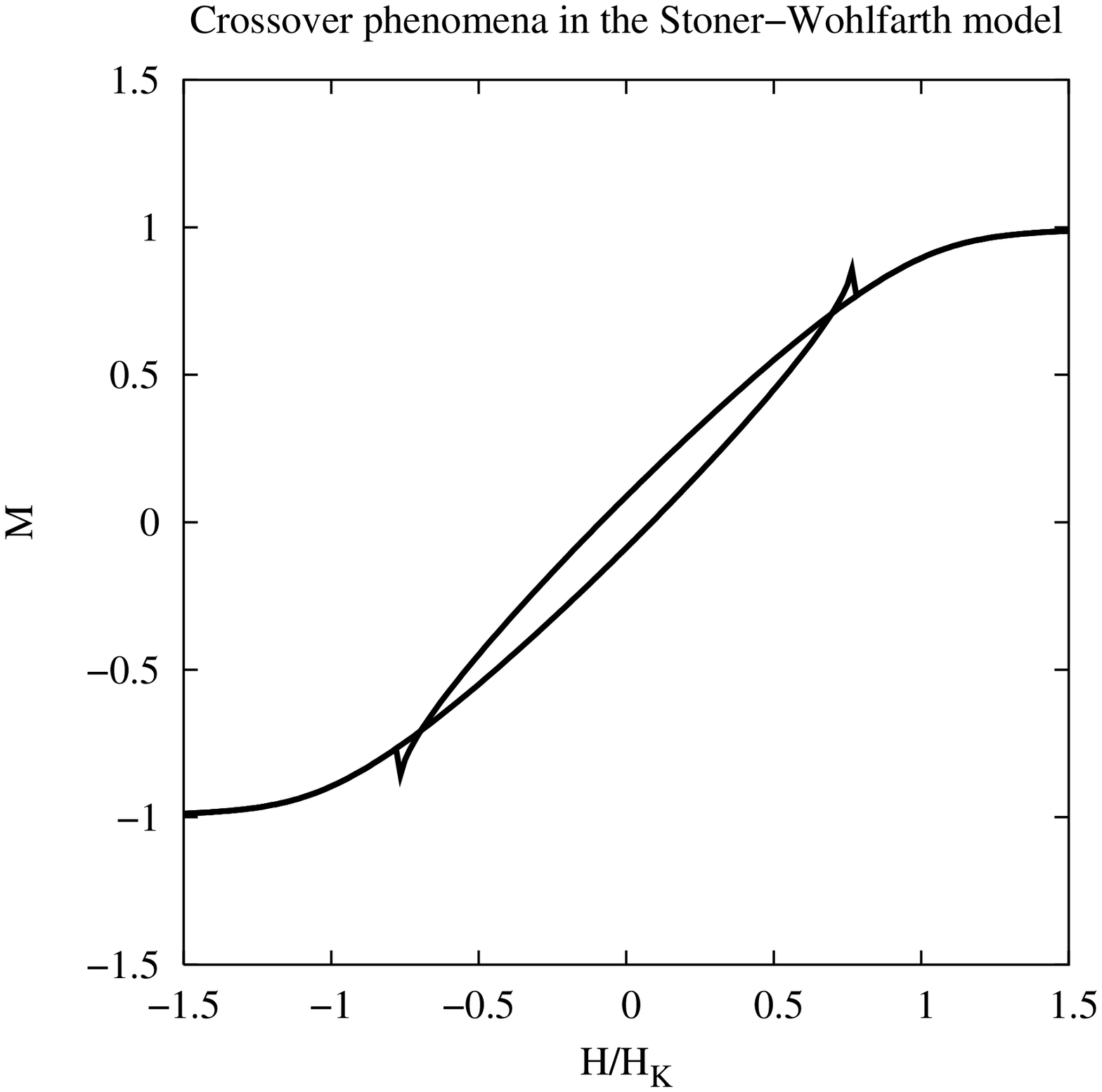}}
\end{center}
  \caption{Hysteresis loop showing the crossover effect for an angle $\phi$=85 \deg}
\label{cross}
\end{figure}

For large $\phi$ angles (typically $>$ 76 \deg) the hysteresis branches 
cross as seen in fig.~\ref{cross} for the particular case $\phi=85$ \deg.

The branch crossing problem has been observed in the SW work but no remedy was offered.
One may slightly modify the expression of the branches for large $\phi$ angles as done
in the work of Stancu and Chiorescu \cite{stancu}.

Another limitation is the ellipsoidal form of the grain used to simplify
demagnetization fields. 

The uniaxial anisotropy of the grain appeals to materials such as Cobalt, whereas
other transition metal FM's such as Iron or Nickel possess cubic anisotropy.
Other types of anisotropy occur as discussed in the second part of this work. \\

Several other limitations such as overestimation of the coercive field, linear dependence
with respect of the energy with respect to grain volume and quantum effects (see for instance
the review by Awschalom and DiVincenzo \cite{awsch}). 

Despite all these limitations, the SW provide a correct overall picture in 
uniformly magnetised materials even if some inadequacies exist in explaining certain 
experimental details. \\

{\bf Acknowledgement} \\
The  authors wish to acknowledge friendly discussions with M. Cormier (Orsay)
regarding dynamic effects in the SW model, N. Bertram (San Diego) and M. Acharyya (U. K\"oln) for sending 
some of their papers prior to publication.\\

\begin{center}
{\bf FIGURES}
\end{center}

\begin{thebibliography}{99}
\bibitem{energy} All energies are taken per unit volume of the particle.
\bibitem{ea} The anisotropy axis is intrinsic and determined by the growth conditions and
cristallography of the grain, whereas the EA is determined from all contributing
energy terms (anisotropy, shape, demagnetization and mutual interactions when they
occur between grains) except Zeeman, since it is the probe. 
The total energy minimum is achieved when the magnetization is along the resulting EA orientation.
\bibitem{note} Anisotropy energy can be considered a contribution to energy given by 
the term $K\sin^{2}\theta$ where $\theta$ is the angle, the magnetization makes with
the $z$ axis. If $K>0$  the energy minimum (=0) occurs when $\theta=0$ that is alignment 
with the $z$ axis (EA case). Taking the  $z$ axis perpendicular to a thin film or a disk plane ($xy$ plane), 
and considering $K<0$, the energy minimum ($-|K|$) is obtained for $\theta=\pi/2$ favoring an in-plane 
magnetization (easy plane case).
\bibitem{jump} There exist cases where higher derivatives of the energy versus angle are
needed to be nulled or that several energy landscapes with the magnetization
taking the value for which the two landscapes cross (see for instance Bertotti's~\cite{Bertotti} book).
\end{thebibliography}
\end{document}